\documentclass[aps,pra,10pt,superscriptaddress,floatfix,reprint]{revtex4-2}
\usepackage{graphicx,bm,dsfont,amsmath,amsthm,amssymb,mathrsfs,hyperref,mdframed,xcolor,soul}
\usepackage{orcidlink}
\newcommand{\ket}[1]{\left| #1 \right\rangle}

\newcommand{\braket}[2]{\langle #1|#2 \rangle}

\begin{document}

\title{Deterministic Teleportation and Universal Computation Without Particle Exchange}

\author{Hatim Salih}
\affiliation{School of Physics, Engineering and Technology, University of York, Heslington, York, YO10 5DD, UK}
\author{Jonte R. Hance\,\orcidlink{0000-0001-8587-7618}}
\email{jonte.hance@newcastle.ac.uk}
\affiliation{School of Computing, Newcastle University, 1 Science Square, Newcastle upon Tyne, NE4 5TG, UK}
\affiliation{Quantum Engineering Technology Laboratory, Department of Electrical and Electronic Engineering, University of Bristol, Woodland Road, Bristol, BS8 1UB, UK}
\author{Will McCutcheon}
\affiliation{Institute of Photonics and Quantum Science, School of Engineering and Physical Sciences, Heriot-Watt University, Edinburgh, EH14 4AS, UK}
\author{Terry Rudolph}
\affiliation{Department of Physics, Imperial College London, Prince Consort Road, London SW7 2AZ, United Kingdom}
\author{John Rarity}
\affiliation{Quantum Engineering Technology Laboratory, Department of Electrical and Electronic Engineering, University of Bristol, Woodland Road, Bristol, BS8 1UB, UK}

\date{\today}

\begin{abstract}
Teleportation is a cornerstone of quantum technologies, and has played a key role in the development of quantum information theory. Pushing the limits of teleportation is therefore of particular importance. Here, we apply a different aspect of quantumness to teleportation---namely exchange-free, or counterfactual, computation at a distance. We propose a universal controlled-phase gate, where no particles are exchanged between control and target. This allows the full repertoire of quantum computation to me made exchange-free, including both complete Bell detection among two remote parties and so teleportation and telecloning. Further, we show that this gate, and the protocols based on it, is experimentally feasible, simulating the fidelity of our exchange-free teleportation and telecloning protocols for realistic {bit-errors}.
\end{abstract}

\maketitle

\section{Introduction}
In the popular imagination, teleportation has come to refer to the process by which a body or an object is transported from one place to another without taking the actual journey. While a staple of science fiction, science by contrast seemed to rule it out based on the uncertainty principle, which placed a fundamental limit on the accuracy of measurement \cite{Kennard1927Quantenmechanik}. No wonder when in 1993 Bennett and colleagues proposed the first quantum teleportation protocol \cite{Bennett1993Teleporting}, it was soon recognised as a seminal moment in physics. Relying on the non-classical resource of pre-shared entanglement between the communicating parties, an unknown quantum state of a physical system is jointly measured by the sender with one part of an entangled pair, in such a way that allows its reconstruction at the remote entangled partner at the receiver, while leaving behind its physical constituents. Classical communication is typically required to complete this disembodied transport. 

Not only has quantum teleportation become a backbone of quantum technologies such as quantum communication, quantum computing, and quantum networks, it has also played a crucial role in the development of formal quantum information theory \mbox{\cite{Horodecki2021QIReview,Flamini2018PhotonicQIP,Kok2010IntroOptQIP,nielsen_chuang_2010}}. As such, pushing the limits of quantum teleportation is of significant importance, which is what we intend to do here by invoking yet another aspect of quantumness: exchange-free computation at a distance. While Gedanken, or thought, experiments have historically played a crucial conceptual role in physics (with the EPR proposal for instance being famously conceived as such) we try to go beyond theory to proposing feasible demonstration.

In exchange-free communication, also known as counterfactual communication, a classical message is sent by means of quantum processes without the communicating parties exchanging any particles. (We use the term exchange-free in place of counterfactual since the term counterfactual has historically been used differently in the literature. Moreover, the term exchange-free more accurately describes the protocol, namely information being sent without exchange of particles.) With its roots in the phenomena of interaction-free measurement and the quantum Zeno effect \cite{Elitzur1993Bomb,Kwiat1995IFM,Kwiat1999Interrogation,Rudolph2000Zeno,Mitchison2001CFComputation,Hosten2006CounterComp,Hance2021CFGI}, the first such deterministic protocol was proposed by Salih et al \cite{Salih2013Protocol}, before being experimentally demonstrated by Pan and colleagues \cite{Pan2017Experiment}. While once controversial, the debate over whether exchange-free communication was permitted by the laws of physics (for both bit values) seems now to be resolving; Nature does allow exchange-free communication, and consequently computation at a distance \cite{Vaidman2014SalihCommProtocol,Salih2014ReplyVaidmanComment,Griffith2016Path,Salih2018CommentPath,Griffiths2018Reply,Aharonov2019Modification,Salih2018Laws,Hance2021Quantum}.

This counterfactual communication was generalised to sending quantum information exchange-free for the first time in the Salih14 protocol \cite{Salih2016Qubit,*Salih2014Qubit}, also known as counterportation, proposing an exchange-free quantum CNOT gate as a new computing primitive. The exchange-free CNOT was later employed by Zaman et al to propose exchange-free Bell analysis, albeit with a 50\% theoretical efficiency limit \cite{Zaman2018SalihBell}.

Significantly, the controlled $\hat{R}_z$-rotation we propose here, based on the above-mentioned CNOT gate, is both universal and has no theoretical limit on efficiency. By contrast, the gates proposed in \mbox{\cite{Salih2016Qubit,*Salih2014Qubit,guo2014counterfactual}} have a theoretical limit on efficiency in the absence of classical communication, while the one in \mbox{\cite{Li2019CFStateEx}}, besides being impractically complicated, is not universal. We then combine quantum teleportation with exchange-free computation at a distance to propose deterministic exchange-free teleportation. The core of this gate is set up by the entangling operation enabled by a one-dimensional cavity atom system. The ground state of an atom in a cavity can be put into a superposition of being on-resonant (a zero) and reflecting, or off-resonant (a one) and transmitting, a photon \cite{Hu2009QuantumDot,Reiserer2015Cavity}. We then construct a counterfactual way of probing whether Bob is blocking/not blocking (transmitting/reflecting) using the standard counterfactual communication-style protocol.

\section{Universal Controlled-Z Rotation}
\begin{figure*}
    \centering
    \includegraphics[width=\linewidth]{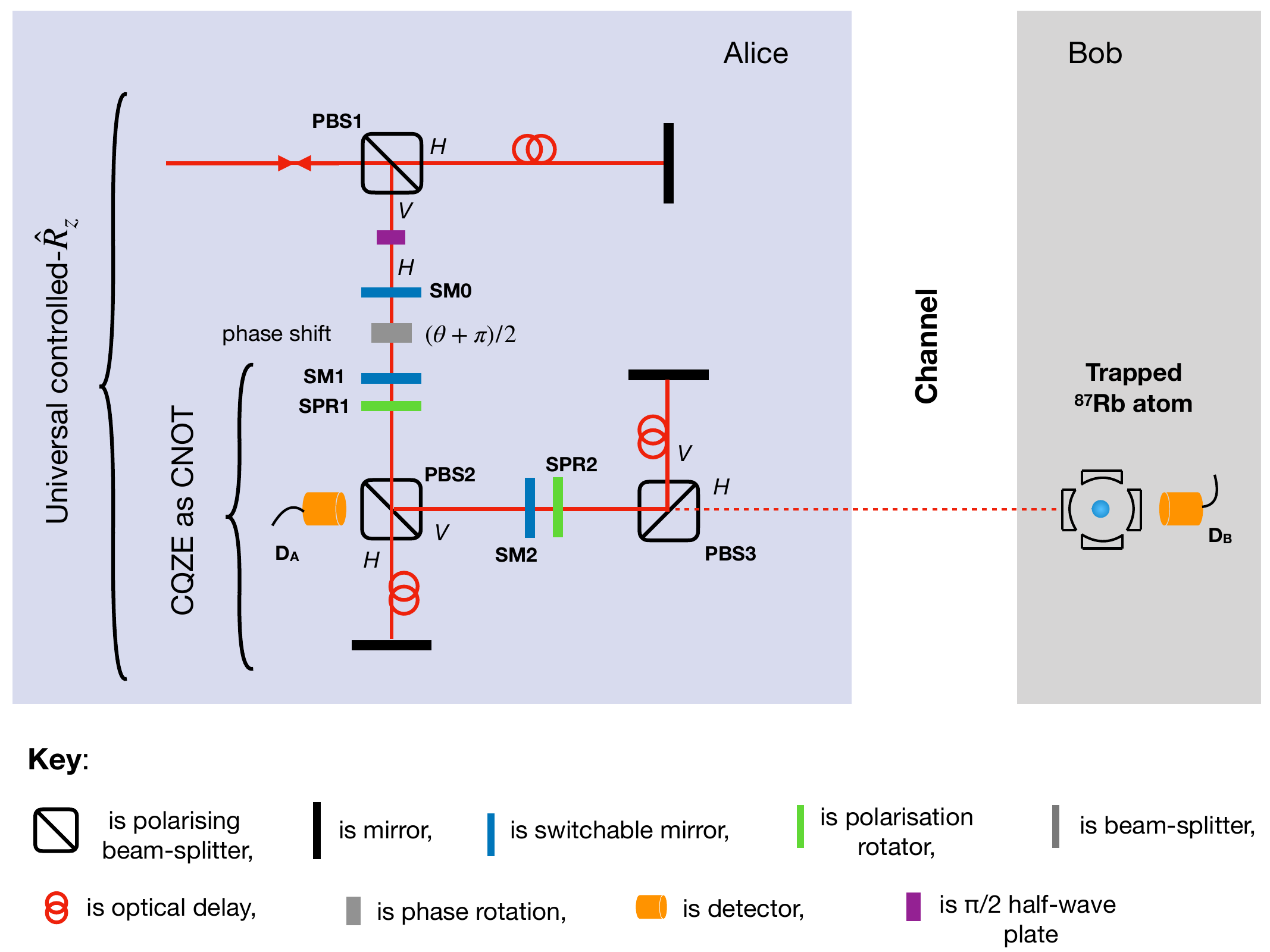}
    \caption{Our setup for an experimentally feasible, exchange-free controlled-$\hat{R}_z$, universal gate. This is based on Salih's exchange-free CNOT gate which has Bob enacting a superposition of blocking and not blocking the communication channel by means of a trapped atom \cite{Salih2016Qubit, *Salih2014Qubit, Salih2018Paradox}. With the addition of a phase-shift plate applying a $(\theta+\pi)/2$ rotation, switchable mirror SM0, a $\pi/2$ half wave plate that flips polarisation, polarising beamsplitter PBS1, and an optical delay loop, the chained quantum Zeno effect unit becomes the basis of our controlled phase-rotation universal gate, entangling the states of Alice's photonic qubit and Bob's trapped atom qubit.}
    \label{fig:CZGate}
\end{figure*}

\subsection{\texorpdfstring{{Chained Quantum Zeno Effect Unit}}{Chained Quantum Zeno Effect Unit}}
We first go through the chained quantum Zeno effect (CQZE) unit, as given in Fig.~\ref{fig:CZGate}. This is based on Salih's exchange-free CNOT gate, which has Bob enacting a superposition of blocking and not blocking his side of the communication channel \cite{Salih2016Qubit,*Salih2014Qubit, Salih2018Paradox}. 
The switchable mirror, SM1, is first switched off to allow the photon into the outer interferometer, before being switched on again. The switchable polarisation rotator, SPR1, rotates the photon's polarisation from $H$ {towards} $V$, by a small angle $\frac{\pi}{2M}$: 
\begin{equation}
\begin{split}
    \left| H \right\rangle \to \cos \frac{\pi}{2M} \left| H \right\rangle + \sin \frac{\pi}{2M} \left| V \right\rangle\\
    \left| V \right\rangle \to \cos \frac{\pi}{2M} \left| V \right\rangle - \sin \frac{\pi}{2M} \left| H \right\rangle
    \end{split}
\end{equation}

The polarising beam-splitter, PBS2, passes the $H$-polarised component towards the mirror below it, while reflecting the small $V$-polarised component towards the inner interferometer. The switchable mirror, SM2, is then switched off to allow the $V$-polarised component into the inner interferometer, before being switched on again. The switchable polarisation rotator, SPR2, rotates the $V$-polarised component by a small angle $\frac{\pi}{2N}$:
\begin{equation}
    \left| V \right\rangle \to \cos \frac{\pi}{2N} \left| V \right\rangle - \sin \frac{\pi}{2N} \left| H \right\rangle
\end{equation}

The polarising beamsplitter, PBS3, then reflects the $V$-polarised component up, towards a mirror while passing the $H$-polarised component towards Bob's trapped atom, which is in a superposition, $\alpha \left| 0 \right\rangle + \beta \left| 1 \right\rangle$, of reflecting back any photon, and transmitting it through to the loss detector. If the atom reflects (is in state $\ket{0}$), the $H$-polarised component survives---if the atom transmits ($\ket{1}$), the component is lost.  This means the overall action of the inner interferometer (assuming the photon is not lost to Bob's detector $D_B$) is

\begin{equation}
\begin{split}
\ket{V}& (\alpha \ket{0} + \beta \ket{1})\\
\to&
\alpha (\cos \frac{\pi}{2N} \ket{V} - \sin \frac{\pi}{2N} \ket{H})\ket{0}\\
&+ \beta \cos \frac{\pi}{2N} \ket{V} \ket{1}
\end{split}
\end{equation}

After N such cycles the photonic superposition has now been brought back together by PBS3 towards SM2. Because the $\ket{0}$ term coherently rotates from $\ket{V}$ to $\ket{H}$, after $N$ such cycles the state is,
\begin{equation}
\begin{split}\label{eqNinner}
\ket{V} (\alpha \ket{0} + \beta \ket{1}) \to \alpha \ket{H} \ket{0} + \beta {\cos}^N \frac{\pi}{2N} \ket{V} \ket{1}
\end{split}
\end{equation}

The switchable mirror, SM2, is then switched off to let the photonic component inside the inner interferometer out. Since for large $N$, ${\cos}^N \frac{\pi}{2N}\rightarrow1$, the implemented transformation becomes,
\begin{equation}
\begin{split}
\ket{V} (\alpha \ket{0} + \beta \ket{1}) \to \alpha \ket{H} \ket{0} + \beta \ket{V} \ket{1}
\end{split}
\end{equation}

Note that this state is entangled, but the $\ket{H}\ket{0}$ term {in this superposition} has been to {and} from Bob, so {using this component for the computation} wouldn't be counterfactual. Consequently, we send this term to loss channel  $D_A$ via PBS2.

The first outer {interferometer}, starting with the photon at SM1 (assuming the photon is neither lost to to Alice's detector $D_A$ nor to Bob's $D_B$) implements first {the switchable polarisation rotator} SPR1, then $N$ inner cycles, {meaning it implements the transformation}
\begin{equation}\label{eqoneouter}
\begin{split}
    \ket{H}& (\alpha \ket{0} + \beta \ket{1})\\
    \to& \left(\cos\frac{\pi}{2M}\ket{H} + \sin\frac{\pi}{2M}\ket{V}\right) (\alpha \ket{0} + \beta \ket{1})\\
    \to&\cos\frac{\pi}{2M}\ket{H}\left(\alpha\ket{0}+\beta\ket{1}\right)\\
    &+ \sin \frac{\pi}{2M} \ket{V}\cos^N\frac{\pi}{2N} \ket{1}
    \\=&\alpha \cos \frac{\pi}{2M} \ket{H} \ket{0} + \beta (\cos \frac{\pi}{2M} \ket{H}\\
    &+ \cos^N\frac{\pi}{2N}\sin \frac{\pi}{2M} \ket{V}) \ket{1}
\end{split}
\end{equation}
{This is} as, after the initial SPR1 rotation, the $H$-polarised component goes {into a} delay {line}, while the $V$-polarised component goes {through} $N$ inner {interferometers}, and so evolves as per Eq.~\ref{eqNinner}, with the $\ket{H}\ket{0}$ component generated by this sent to loss mode $D_A$.

This represents one outer cycle, containing $N$ inner cycles. The photonic superposition has now been brought back together by PBS2 towards SM1. After $M$ such cycles (setting ${\cos}^N \frac{\pi}{2N}\rightarrow1$), the coherent rotation of the $\ket{1}$ part of the state from $\ket{H}$ to $\ket{V}$ means the protocol applies
\begin{equation}\begin{split}\label{eqMouter}
\ket{H} (\alpha \ket{0} + \beta \ket{1}) \to \alpha {\cos}^M \frac{\pi}{2M} \ket{H} \ket{0} + \beta \ket{V} \ket{1}
\end{split}\end{equation}

Since for large $M$, ${\cos}^M \frac{\pi}{2M}$ approaches 1, Eq.~\ref{eqMouter} goes to,
\begin{equation}\begin{split}
\ket{H} (\alpha \ket{0} + \beta \ket{1}) \to \alpha \ket{H} \ket{0} + \beta \ket{V} \ket{1}
\end{split}\end{equation}

The switchable mirror SM1 is now switched off to let the photon out {of the outer interferometer}. Crucially, this last equation describes the action of a quantum CNOT gate with Bob's as the control qubit, acting on Alice's $H$-polarised photon.

{Rather than considering} the large-$M$ limit, let {us instead} illustrate {the protocol} for $M=2$. Taking the state at the end of Eq.~\ref{eqoneouter}, setting $\cos^N\frac{\pi}{2N}\to1$, and putting it through SPR1 again, we get
\begin{equation}
    \begin{split}
        \cos& \frac{\pi}{2M}\ket{H}\left(\alpha\ket{0}+\beta\ket{1}\right) + \sin \frac{\pi}{2M} \ket{V} \ket{1}\\
        \to & \cos\frac{\pi}{2M}\left(\cos\frac{\pi}{2M}\ket{H} + \sin\frac{\pi}{2M}\ket{V}\right)\left(\alpha\ket{0}+\beta\ket{1}\right)\\
        &+ \sin \frac{\pi}{2M} \left(\cos\frac{\pi}{2M}\ket{V} - \sin\frac{\pi}{2M}\ket{H}\right) \ket{1}\\
        =&_{\text{For }M=2} \frac{1}{\sqrt{2}}\left(\frac{1}{\sqrt{2}}\ket{H} + \frac{1}{\sqrt{2}}\ket{V}\right)\left(\alpha\ket{0}+\beta\ket{1}\right)\\
        &+ \frac{1}{\sqrt{2}}\left(\frac{1}{\sqrt{2}}\ket{V} - \frac{1}{\sqrt{2}}\ket{H}\right) \ket{1}\\
        =&\frac{\ket{H}}{2}\alpha\ket{0} + \ket{V}\left(\frac{\alpha}{2}\ket{0}+\beta\ket{1}\right)
    \end{split}
\end{equation}

If we now send this through $N$ more inner interferometers, due to the $\ket{H}$ component being kept in the delay below PBS2, and the $\ket{V}\ket{0}$ component coherently rotating to $\ket{H}\ket{0}$ and so being lost to loss channel $D_A$, we get the transformation for the entire two outer-cycle ($M=2$) protocol as 
\begin{equation}
\begin{split}
    \ket{H} (\alpha \ket{0} + \beta \ket{1}) \to \frac{\ket{H}}{2}\alpha\ket{0} + \ket{V}\beta\ket{1}
\end{split}
\end{equation}

\subsection{\texorpdfstring{{Exchange-Free Controlled-Phase Gate}}{Exchange-Free Controlled-Phase Gate}}
We now explain the rest of the setup, which uses the CQZE unit to implement a universal, general-input controlled-$\hat{R}_z$ rotation gate (Fig.~\ref{fig:CZGate}). We begin with a superposition state at Alice, $a\ket{V}+b\ket{H}$. This is split at PBS1, with the $H$-polarised component going into an optical loop, and the $V$-polarised component going through a $\pi/2$ half wave plate flipping its polarisation to $H$, before being admitted into the CQZE unit by turning off the switchable mirror SM0, before turning it on again. Upon exiting the CQZE unit, it is reflected back by SM0, having a phase of $\theta+\pi$ if $V$-polarised (0 if $H$) applied to it by the phase shifter, before going through another run of the CQZE unit. This always produces an $H$-polarised state, as noted in \cite{Li2019CFStateEx}. Note that the $\pi$ term in the phase shifter is a correction term. The photonic component now exits through SM0, which is switched off, before being flipping back to $V$-polarisation at the $\pi/2$ half wave plate, having acquired a $\theta$ phase shift. It then recombines with the $H$-polarised component at PBS1.

Given the initial state of the overall system is 
\begin{equation}
    (a\ket{V}+b\ket{H})\otimes(\alpha\ket{0}+\beta\ket{1})
\end{equation}
and that only the initially $V$-polarised component ``interacts" with the trapped atom, we get the final state
\begin{equation}
    a\ket{V}(\alpha\ket{0}+\beta e^{i\theta}\ket{1})+b\ket{H}(\alpha\ket{0}+\beta\ket{1})
\end{equation}

This is an entangled state between Alice's polarisation qubit and Bob's trapped ion qubit: a controlled-phase rotation, with Alice's as the control qubit and Bob's as the target. Due to the symmetry of control and target qubits for this type of rotation, it can also be factorised as
\begin{equation}
     \alpha\ket{0}(a\ket{V}+b\ket{H})+\beta\ket{1}(ae^{i\theta}\ket{V}+b\ket{H})
\end{equation}
the same controlled-phase rotation expressed differently, now with Bob's as the control qubit and Alice's as the target. Taking the special case when $\theta=\pi$, we get a controlled-Z gate.

{Note that this} exchange-free controlled-$\hat{R}_z$ {gate} allows efficient implementation of any quantum circuit when combined with local operations {- together these form a universal set}. But there's another sense in which it {this counterfactual controlled-phase rotation is} universal. As explained later, this gate can be operated differently, allowing one party with classical action to enact any desired operation on a second party's remote photonic qubit, exchange-free. This classical action can even control a two-qubit gate at the second party, as shown in \cite{Salih2021EFQubit}. Our controlled-$\hat{R}_z$ gate can therefore be thought of as a universal set in its own right. {Given a universal set of quantum gates can be used to perform any possible quantum computation, this means the exchange-free controlled-Z gate in this paper allows us to perform any possible quantum computation or quantum algorithm, exchange free. We later use this to demonstrate two key protocols---exchange-free teleportation, and exchange-free telecloning.}

\subsection{\texorpdfstring{{Experimental Implementation using Trapped Atoms}}{Experimental Implementation using Trapped Atoms}}

{In order to implement this gate,} Bob needs a way to {not only reflect} (bit ``0''), and {block} (bit ``1''), Alice's photon, {but to do superpositions of this reflecting and blocking}. Recent demonstration{s} of light-matter quantum logic gates {using} trapped atoms inside optical cavities \cite{Reiserer2015Cavity,Tiecke2014NanophotonicAtom,Reiserer2014a}, {suggest} trapped atoms {are an ideal way to implement this}.

{We can therefore implement} Bob's qubit {by using an} $^{87}$Rb atom trapped {within} a high-finesse optical {resonant cavity using} a three-dimensional optical lattice \cite{Reiserer2014a,Reiserer2014}. {This $^{87}$Rb atom has two internal ground states---}depending on which of {these} the atom is in, a photon {arriving at} the cavity in Fig.~\ref{fig:CZGate} either {strongly couples to the atom, and so is} reflected, or {travels through the} cavity {and enters} detector $D_B$. {This requires the cavity to have} mirror reflectivities such that photon{s} entering the cavity {exit} towards detector $D_B$ ({as in Ref.~\mbox{\cite{Mucke2010}})}. {Raman transitions, implemented using paired Raman lasers, allow us to place the} $^{87}$Rb {atom} in a superposition of {these} two ground states. {This allows} Bob {to} implement {his} desired superposition of reflecting {and absorbing} Alice's photon. {The} coherence time for {Ref.~\mbox{\cite{Reiserer2014}}'s implementation of such a superposition in a trapped-atom} system is of {order} $10^{-4}$ {seconds}. Therefore, if the protocol is completed within $\mathcal{O}(10^{-5}s)$, {we can ignore} decoherence effects {for this system}. {The key factor lower-bounding this is that it requires a} $\mathcal{O}(10^{-9}s)$ switching speed {for the} switchable components, {which is not currently experimentally feasible}. {However, this reliance on} switchable optical elements {could also be mitigated by using} experimental trick, as {was done in} Cao et al's experimental implementation of Salih et al's 2013 protocol \cite{Pan2017Experiment}.

\section{Exchange-free Teleportation}

\begin{figure*}
    \centering
    \includegraphics[width=\linewidth]{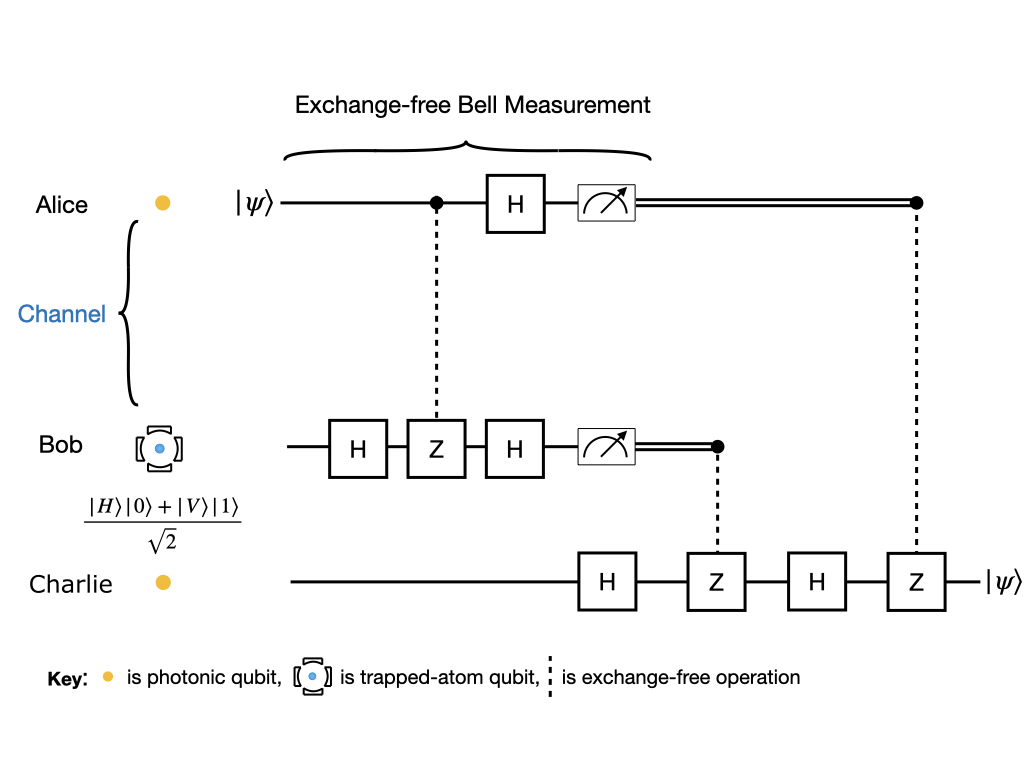}
    \caption{Our protocol for exchange-free teleportation, sending a quantum state from Alice via Bob to Charlie. In this protocol, Alice has a photon-polarisation qubit, while Bob and Charlie have a maximally-entangled pair of qubits, Bob's implemented as a trapped-atom enacting a superposition of blocking and not blocking the communication channel, and Charlie's as photon polarisation. Alice's qubit begins in the state to be teleported, $\ket{\psi}$. Importantly, complete Bell detection takes place without Alice and Bob (or Alice and Charlie, or Bob and Charlie) exchanging any particles, and instead of classical communication, Alice and Bob can directly apply X and Z operations on Charlie's photonic qubit as required based on their respective measurements. The three exchange-free Controlled-Z gates are instances of the set-up of Fig.~\ref{fig:CZGate}. Similarly, Bob and Charlie's initial entanglement can also be set up using an exchange-free Controlled-Z gate and three Hadamard gates (one before the Controlled-Z gate for both Bob and Charlie's qubits, and one after for Bob's qubits, if the qubits are initialised in states $\ket{0}$ and $\ket{H}$). This means, even if we partition the teleportation in the normal way (input qubit and one part of the entangled state at Alice-Prime, and second part of the entangled state at Bob-Prime), the protocol can still occur entirely counterfactually.}
    \label{fig:Teleprot}
\end{figure*}

We now {show how we can use this exchange-free controlled-Z gate to form the basis of an} exchange-free implementation of teleportation. This is based on the quantum teleportation first devised by Bennett \textit{et al} \cite{Bennett1993Teleporting}, but recast such that there is no need for previously-shared entanglement between the two communicants (here, Alice and Charlie), nor classical non-counterfactual communication between then. Our {exchange-free} teleportation scheme is shown in Fig.~\ref{fig:Teleprot}. 

In this protocol, we have a photon-polarisation qubit at Alice, and an entangled pair of qubits, one trapped-atom and the other photon-polarisation, at Bob and Charlie respectively. Alice's qubit is instantiated in the state to be teleported, e.g. by a third party, while Bob and Charlie's two modes are in the maximally entangled state 
    $\ket{H}\ket{0}+\ket{V}\ket{1}/\sqrt{2}$
 (which can be created exchange-free using an exchange-free CZ-gate and two Hadamard gates, to enact an exchange-free CNOT, plus an extra Hadamard gate).

To enact teleportation, Bob first applies a Hadamard gate to his trapped-atom qubit, before Alice applies an exchange-free controlled-Z gate, with her photonic qubit as the control and and Bob's trapped-atom qubit as the target. Bob and Alice then apply Hadamard gates onto their respective qubits, before measuring the states in the computational basis for Bob, and in the $H/V$ basis for Alice, together performing a complete Bell measurement. Bob then either flips or doesn't flip the polarisation of Charlie's photonic qubit based on the classical measurement outcome of his trapped-atom qubit. He does this as follows. Based on the classical measurement outcome of his qubit, he either performs an exchange-free controlled-Z on Charlie's photonic qubit with his control set to $\ket{1}$, or else sets his control to $\ket{0}$. The two Hadamard gates transform the Z operation into an X. Alice then, based on the classical measurement outcome of her qubit, either performs an exchange-free controlled-Z on Charlie's photonic qubit with her control set to $\ket{1}$, or else sets her control to $\ket{0}$. These last two steps by Bob and Alice respectively act as the feed-forward step of teleportation, leaving Charlie's photonic qubit in the state of Alice's original qubit.

\begin{figure}
    \centering
    \includegraphics[width=\linewidth]{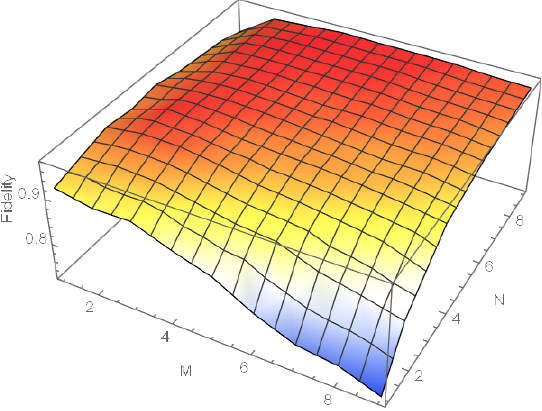}
    \caption{Average fidelity of our exchange-free teleportation protocol as a function of the number of outer and inner cycles, M and N. This is for an imperfect trapped-atom at Bob that fails to reflect an incident photon 34\% of the time when it should reflect, and fails to block the photon 8\% of the time when it should block. Fidelity is averaged over 100 points evenly distributed over possible states Alice could send.}
    \label{fig:Fid}
\end{figure}

This exchange-free protocol bears all the hallmarks of teleportation as given by Pirandola et al \cite{Pirandola2015AdvsTele}. Alice's input state is unknown to her, and can be provided by a third party who also verifies the teleported state at Charlie. The protocol allows complete Bell detection, which in our case is jointly carried out by Alice and Bob exchange-free. The protocol allows the possibility of real-time correction on Charlie's photonic qubit. Lastly, even for the smallest number of cycles, achievable fidelity for our protocol exceeds the 2/3 limit of ``classical teleportation", which comes from the no-cloning theorem \cite{Wootters1982NoCloning}. Fig.~\ref{fig:Fid} gives the average fidelity $F(\theta,\phi)$, where
\begin{equation}\label{eq.fid}
    F(\theta,\phi) = \braket{\Psi_{in}}{\Psi_{out}}\braket{\Psi_{out}}{\Psi_{in}}
\end{equation}
and $\Psi_{in}$ and $\Psi_{out}$ are the input and output states of the protocol, $\theta$ and $\phi$ parameterise the input state's Bloch sphere (azimuthal and radial angle respectively), and the average fidelity is $ F(\theta,\phi)$ averaged over $\theta$ and $\phi$ \cite{Oh2002TeleFid}. The average efficiency of the protocol is 30\% for a number of cycles $M$ equal to 10 and $N$ equal to 100, but improves for larger numbers of cycles.

\section{Telecloning}
\begin{figure}
    \centering
    \includegraphics[width=\linewidth]{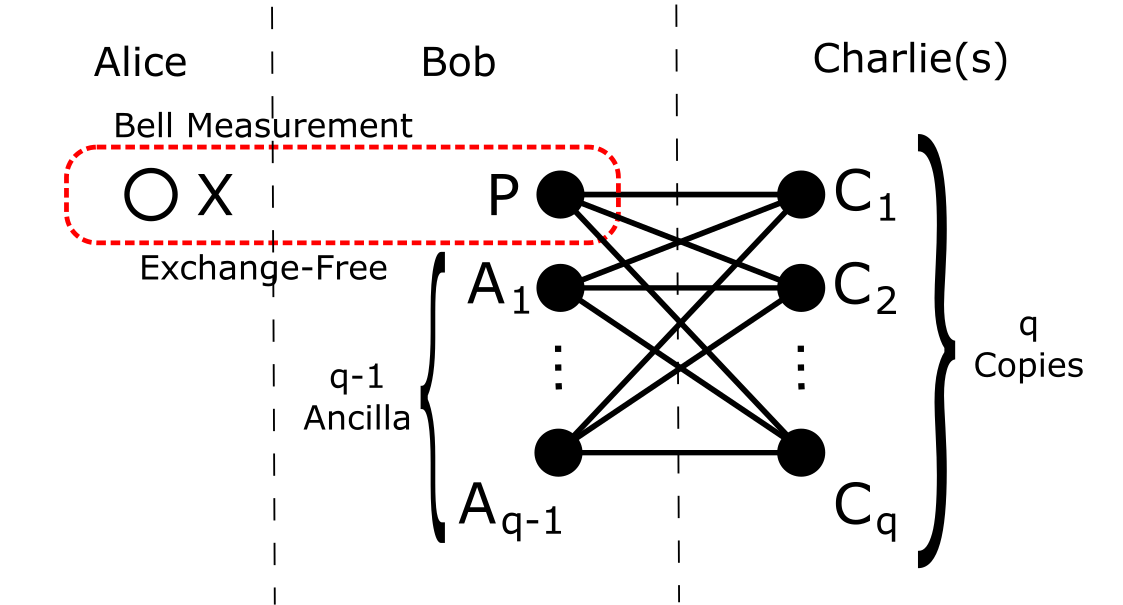}
    \caption{An entanglement diagram for exchange-free telecloning. The protocol starts with an initial entangled state between port-qubit P and ancilla-qubits A$_{q-1}$ at Bob, and the copy-qubits C$_q$ at Charlie(s). If desired, this can be generated exchange-free using combinations of Hadamard gates and exchange-free Controlled-Z gates (as we give above). Alice and Bob perform a Bell Measurement on Bob's port qubit P and Alice's initial state qubit X. The thick black lines mark entanglement, while the red dashed box indicates an exchange-free Bell measurement. This forces the system into one of four states. Alice applies suitable exchange-free controlled-rotations (Pauli operations) on the copy-qubits C$_q$ to recover the approximate copies at Charlie(s).}
    \label{fig:Telecloning}
\end{figure}

\begin{figure}
    \centering
    \includegraphics[width=\linewidth]{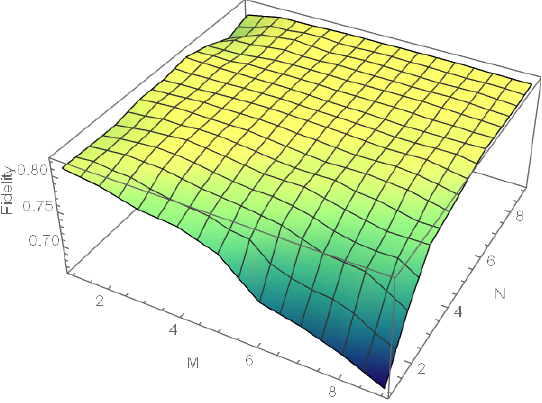}
    \caption{Average fidelity for exchange-free telecloning using the exchange-free controlled-Z gate described above, plotted for different numbers of outer ($M$) and inner ($N$) cycles. Fidelity is calculated for an imperfect trapped-atom at Bob that fails to reflect an incident photon 34\% of the time when it should reflect, and fails to block the photon 8\% of the time when it should block. Fidelity is averaged over 100 points evenly distributed on the Bloch sphere of possible states Alice could send.}
    \label{fig:avgFidTC}
\end{figure}

Since quantum telecloning combines approximate cloning with teleportation to transport multiple approximate copies of a states, one would think that our exchange-free teleportation protocol might allow telecloning to be carried out exchange-free. We can make Murao et al's telecloning scheme \cite{Murao1999Telecloning} exchange-free in a similar manner to how we made teleportation exchange-free, given it also employs a Bell measurement, along with local operations at the receiver based on the Bell detections. Their scheme starts with an already prepared multipartite entangled state \cite{Murao1999Telecloning}, which for our purposes we take to be located with port- and ancilla-qubits at Bob, and copy-qubits at Charlie. One of these entangled qubits (the port qubit) is a trapped-atom, and the copy qubits where the approximate copies appear are all photonic. Alice and Bob jointly perform an exchange-free Bell measurement between Alice's photonic input qubit, and Bob's trapped-atom `port' qubit, as we show in Fig.~\ref{fig:Telecloning}.
Based on the classical outcomes of the Bell measurement, Alice applies suitable exchange-free controlled-rotations (Pauli operations) to recover the approximate copies at Charlie(s). The fidelity of these copies is limited by the no-cloning theorem to
\begin{equation}
    \gamma=\frac{2q+1}{3q}
\end{equation}
where $q$ is the number of approximate copies of our state that we want to send. For the protocol, when the trapped-atom interaction is ideal, we always reach this limit of fidelity (5/6 for two copies) even for the smallest number of cycles. In Fig.~\ref{fig:avgFidTC}, we give the fidelity (as per Eq.~(\ref{eq.fid})) for an imperfect trapped-atom at Bob that fails to reflect an incident photon 34\% of the time when it should reflect, and fails to block the photon 8\% of the time when it should block, for different values of $M$ and $N$. Average efficiency is 14\% for $M$ equal to 10 and $N$ equal to 100, but improves for larger numbers of cycles.

\section{Exchange-Free Phase Rotation without Time-Binning}
In our recent paper, \textit{Exchange-Free Computation on an Unknown Qubit at a Distance}, we give a protocol that allows Bob to implement any phase on Alice's qubit, exchange-free \cite{Salih2021EFQubit}. This then forms the basis of a device that we called a phase unit, allowing Bob to apply any arbitrary single-qubit unitary to the qubit, exchange-free. However, an issue that phase unit displayed was that the time the photon exited the device was correlated with the phase applied by Bob. While we provided a way for Bob to undo this time-binning after the fact, it is generally desirable to remove it altogether.

By adapting the controlled phase-rotation above, a phase unit can be constructed that doesn't exhibit this time-binning. We use the set-up in Fig.\ref{fig:CZGate}, but instead with a classical Bob either blocking or not-blocking, and with SM0 keeping Alice's photon in the device for 2L runs (rather than 2). Bob sets $\theta$ to $\pi/L$, blocking for $2k$ of the runs and not blocking for $2(L-k)$, in units of 2 runs where he either blocks for both or does not block for both. This allows Bob to set a phase on Alice's photon of $2\pi k/L$. We place three of these devices in series, interspersed with a $-\pi/4$-aligned Quarter Wave Plate, $\hat{\textbf{U}}_{QWP}\hat{\textbf{R}}_x(-\pi/2)$, and its adjoint, $\hat{\textbf{U}}^{\dagger}_{QWP} \hat{\textbf{R}}_x(\pi/2)$, to create a chained-$\hat{R}_z \hat{R}_x \hat{R}_z$a set of rotations, into which any single qubit unitary can be decomposed. Bob can thus apply any arbitrary single-qubit unitary to Alice's qubit---both exchange-free, and without time-binning. This also, as we show in \cite{Salih2021EFQubit}, allows us to classically control of a universal two-qubit gate, which enables Bob to directly enact in principle any desired algorithm on a remote Alice's programmable quantum circuit.

\section{Discussion}

{It is worth noting that, despite being counterfactual, the proposed counterfactual teleportation protocol does in no way potentially enable superluminal communication---given the protocol requires (typically much longer than) the time light would take to travel from Alice to Bob to Charlie, any communication is entirely sub-luminal. Further, for the teleportation to be effective, it still requires the classical (counterfactual) communication in the feed-forward step, meaning the protocol in no way violates the no-signalling condition. This similarly applies for the counterfactual telecloning protocol.}

{Similarly, despite being counterfactual, the counterfactual controlled-phase/controlled-Z gate is still a quantum gate, rather than falling under the category of ``Local Operations and Classical Communications" (LOCC)---therefore, the fact this counterfactual gate can be used to entangle two systems does not violate any theorems showing that LOCC cannot be used to increase entanglement.}

{Given we require at least two outer interferometers, each of at least two inner interferometers, in order for the exchange-free controlled-phase/Z gate to work, implementing such a gate will inevitably take far more time than an equivalent non-exchange-free controlled-phase/controlled-Z gate would take. Similarly, the system we give in Fig.~\mbox{\ref{fig:CZGate}} is far more complex than non-exchange-free version of these gates. However, this is beside the point---the fact that such gates can be made exchange-free, and can be used to form a universal counterfactual gate set, meaning any quantum computation can be performed counterfactually, raises interesting questions about the power of quantum interference beyond cases with ``particle exchange'', and about the nature of quantum computation more generally.}

An interesting modification of Salih et al's 2013 protocol was recently proposed by Aharonov and Vaidman, satisfying their criterion, based on weak trace, for exchange-free communication \cite{Aharonov2019Modification}. Following Salih's 2018 paper on counterportation \cite{Salih2018Paradox}, we now show how to implement it in our protocol. In the CQZE module, after applying SPR2 inside the inner interferometer for the $N$th cycle, Alice now makes a measurement by blocking the entrance to channel leading to Bob. (She may alternatively flip the polarisation and use a PBS to direct the photonic component away from Bob.) Instead of switching SM2 off, it is kept turned on for a duration corresponding to $N$ more inner cycles, after which SM2 is switched off as before. One has to compensate for the added time by means of optical delays. The idea here is that, for the case of Bob not blocking, any lingering $V$ component inside the inner interferometer after $N$ inner cycles (because of weak measurement or otherwise) will be rotated towards $H$ over the extra $N$ inner cycles. This has the effect that, at least as a first order approximation, any weak trace in the channel leading to Bob is made negligibly small.

While alternative proposals have been given for counterfactual communication by Arvidsson-Shukur et al. \cite{Arvidsson2016Communication,Calafell2019Trace}, the protocols' definition of counterfactuality have been the subject of debate \cite{Vaidman2019Analysis,Hance2021Quantum,Wander2021Three}. Their adoption of Fisher Information as a tool for analysing counterfactuality is interesting nonetheless.

As Vaidman points out in \cite{Vaidman2016Counterfactual}, Salih's 2014 protocol---also known as counterportation \cite{Salih2018Paradox}---achieves the same end goal of Bennett et al.'s (1993) teleportation \cite{Bennett1993Teleporting}, namely the disembodied transport of an unknown quantum state, albeit over a large number of protocol cycles. It is an entirely different protocol though, as can be seen from their respective quantum-circuit diagrams. Our current protocol, by contrast to counterportation, is directly based on the 1993 teleportation protocol, but implemented using our universal, newly proposed counterfactual Z-gate, with the aim of exploring the foundations of this most central of quantum information protocols. 

{Future work will aim to simulate the exchange-free controlled phase/controlled-Z gate we give in this paper, and the protocols we give making use of this, for realistic experimental component losses and noise (beyond the 34\% fail-to-reflect and 8\% fail-to-block imperfections we show fidelities for in Figs.~\mbox{\ref{fig:Fid}} and \mbox{\ref{fig:avgFidTC}}), and ideally to work with experimentalists to demonstrate the protocol in practice.}

In conclusion, we have shown how the chained quantum Zeno effect can be employed to construct an experimentally feasible, exchange-free controlled-$\hat{R}_z$ operation, which is not only a universal gate, but can be considered a universal set. This allowed us to propose a protocol for deterministic teleportation of an unknown quantum state between Alice and Bob, without exchanging particles.
\section*{Acknowledgements}

\textit{Funding---}This work was supported by the Engineering and Physical Sciences Research Council (Grants EP/P510269/1, EP/T001011/1, EP/R513386/1, EP/M013472/1 and EP/L024020/1). JRH acknowledges support from a Royal Society Research Grant (RG/R1/251590).\\
\textit{Author Contributions (CRediT)---} HS: Conceptualisation, Formal Analysis, Investigation, Methodology, Software, Visualisation, Writing (Original Draft), Writing (Review and Editing), Supervision, Project Administration; JRH: Formal Analysis, Investigation, Methodology, Software, Visualisation, Writing (Original Draft), Writing (Review and Editing), Project Administration, Supervision; WM: Software, Writing (Review and Editing); TR: Conceptualisation, Investigation, Methodology, Supervision, Writing (Review and Editing); JR: Supervision, Writing (Review and Editing), Funding Acquisition, Resources\\
\textit{Competing Interests---} The authors declare no competing interests.\\
\textit{Data and Materials Availability---} All data needed to evaluate the conclusions in the paper are present in the paper. Code used to reach the conclusions in this paper is available at http://doi.org/10.5281/zenodo.4974506.

\bibliographystyle{apsrev4-2}
\bibliography{Detteleref.bib}

\end{document}